\documentclass{aastex}
\usepackage{spr-astr-addons}
\usepackage{url}
\usepackage{graphics}
\usepackage{epsfig}
\usepackage{longtable}
\RequirePackage{color}
\begin{document}
\title{Revisiting the Amati and Yonetoku Correlations with \textit{Swift} GRBs}
\author{H.~Zitouni}
\affil{Facult\'{e} des sciences, Universit\'{e} Dr Yahia Fares,
P\^{o}le urbain, M\'{e}d\'{e}a, Algeria.}
\email{zitouni.hannachi@gmail.com}\and
\author{N.~Guessoum} \affil{Department of Physics, College of Arts
\& Sciences, American University of Sharjah, UAE.}\email{
nguessoum@aus.edu} \and
\author{W.~J.~Azzam} \affil{Department of Physics, College of
Science, University of Bahrain,
Bahrain.}\email{wjazzam@sci.uob.bh}


\begin{abstract}

We use a sample of \textit{Swift} gamma-ray bursts (GRBs) to
analyze the Amati and Yonetoku correlations. The first relation is
between $E_{p,i}$, the intrinsic peak energy of the prompt GRB
emission, and $E_{iso}$, the equivalent isotropic energy. The
second relation is between $E_{p,i}$  and $L_{iso}$, the isotropic
peak luminosity. We select a  sample of 71 \textit{Swift} GRBs
that have a measured redshift and whose observed  $E^{obs}_p$ is
within the interval of energy 15-150 keV with a relative
uncertainty of less than 70\%. We seek to find correlation
relations for long-duration GRBs (LGRBs) with a peak photon flux
$P_{ph}\geq 2.6~ \mathrm{ph/cm^{2}/s}$. Uncertainties (error bars)
on the values of the calculated energy flux \textit{P}, the energy
$E_{iso}$, and the peak isotropic luminosity $L_{iso}$ are
estimated using a Monte Carlo approach. We find 27 \textit{Swift}
LGRBs that satisfy all our constraints. Results of our analyses of
the  sample of 71 GRBs and the selected subsample (27 GRBs) are in
good agreement with published results. The plots of the two
relations for all bursts show a large dispersion around the best
straight lines in the  sample of 71 LGRBs but not so much in the
subsample of 27 GRBs.
\end{abstract}
\section{Introduction}
 Gamma-ray bursts (GRBs) are sudden,
and very brief, outbursts of high-energy gamma photons that appear
randomly in time and space. They were serendipitously discovered
in 1967, and are of great importance because they are currently
the most luminous and distant sources in the universe. They hold
great promise as cosmological probes of the early universe, since
their flux is unencumbered by extinction due to dust
\citep{{ghirlanda:06}, {azzam:06}, {azzam:06b}, {capozziello:08},
{demianski:11}}. One of the most important elements in the
detection of GRBs is the redshift, \textit{z}, since its
determination is necessary in order to investigate all the
intrinsic characteristics of GRBs. It is generally determined by
the identification of absorption lines in the optical afterglow
spectra, when they are bright enough. Large terrestrial telescopes
equipped with spectrographs working in the infrared or the optical
domains are the best places to perform this task.

Over the past decade, several GRB energy and luminosity relations
have been discovered, in which an observed parameter correlates
with an intrinsic parameter. Some of these relations are obtained
from the light curves, like the lag-luminosity relation
\citep{norris:00} and the variability relation
\citep{fenimore:00}, while others are extracted from the spectra
and include the Amati relation \citep{{amati:02}, {amati:06},
{amati:08}, {amati:09}}, the Ghirlanda relation
\citep{ghirlanda:04}, the Yonetoku relation \citep{{yonetoku:04},
{ghirlanda:10}}, and the Liang-Zhang relation \citep{liang:05}.\\
In this work, we use a sample of \textit{Swift} GRBs that we
selected according to a specific criterion (described in Section
2) in order to investigate two of these correlations: the Amati
relation, which is a correlation between the intrinsic (i.e.,
rest-frame) peak energy, $E_{p,i}$, in a burst's
$\nu\textit{F}_{\nu}$ spectrum and its equivalent isotropic
energy, $E_{iso}$; and the Yonetoku relation which is a
correlation between $E_{p,i}$ and a burst's isotropic peak
luminosity, $L_{iso}$.

A detailed description of our sample selection is provided in
Section 2, which is followed by a presentation of our spectral
analysis and results in Sections 3 and 4, respectively. A
discussion of our results including a comparison with what has
been done by others is given in Section 5, and our conclusions are
provided in Section 6.

\section{Sample Selection}
We use the \textit{Swift} GRBs data that is published on the
official websites
\footnote{http://swift.gsfc.nasa.gov/archive/grb$\_$table/}$^,$\footnote{http://gcn.gsfc.nasa.gov/swift$\_$gnd\_ana.html}.
  The first one presents the observational results
characterizing the overall GRB: peak flux, fluence, duration,
redshift, host galaxy, as well as data on afterglows. The second
website provides more details on the energy spectrum and the time
profile in different energy bands, for different time resolutions.
The data are all provided with their uncertainties (error bars).
We simply use that data to check for the validity of the Amati and
Yonetoku correlations. Bursts that interest us are therefore
long-duration  GRBs (LGRBs) with measured redshifts.

 As of 25/09/2013, \textit{Swift}
observed 809 GRBs, of which 703 are long-duration GRBs. Among
these 703 LGRBs, only 236 have measured redshifts, of which 17 are
`` approximate'' redshifts (060708, 071020, 050904, 110726A,
100704A, 090814A, 070721B, 060912A, \\070306, 050803, 120521C,
060116, 100728B, 081222, 060502A, 080430, and 050802), which
leaves 219 LGRBs with well determined redshifts. We note that in
cases where several methods for the determination of a redshift
were possible, we have adopted the values given by the absorption
method, which is generally the most precise (with four significant
figures).

\begin{figure}
       \centering
       \includegraphics[angle=0, width=0.45\textwidth]{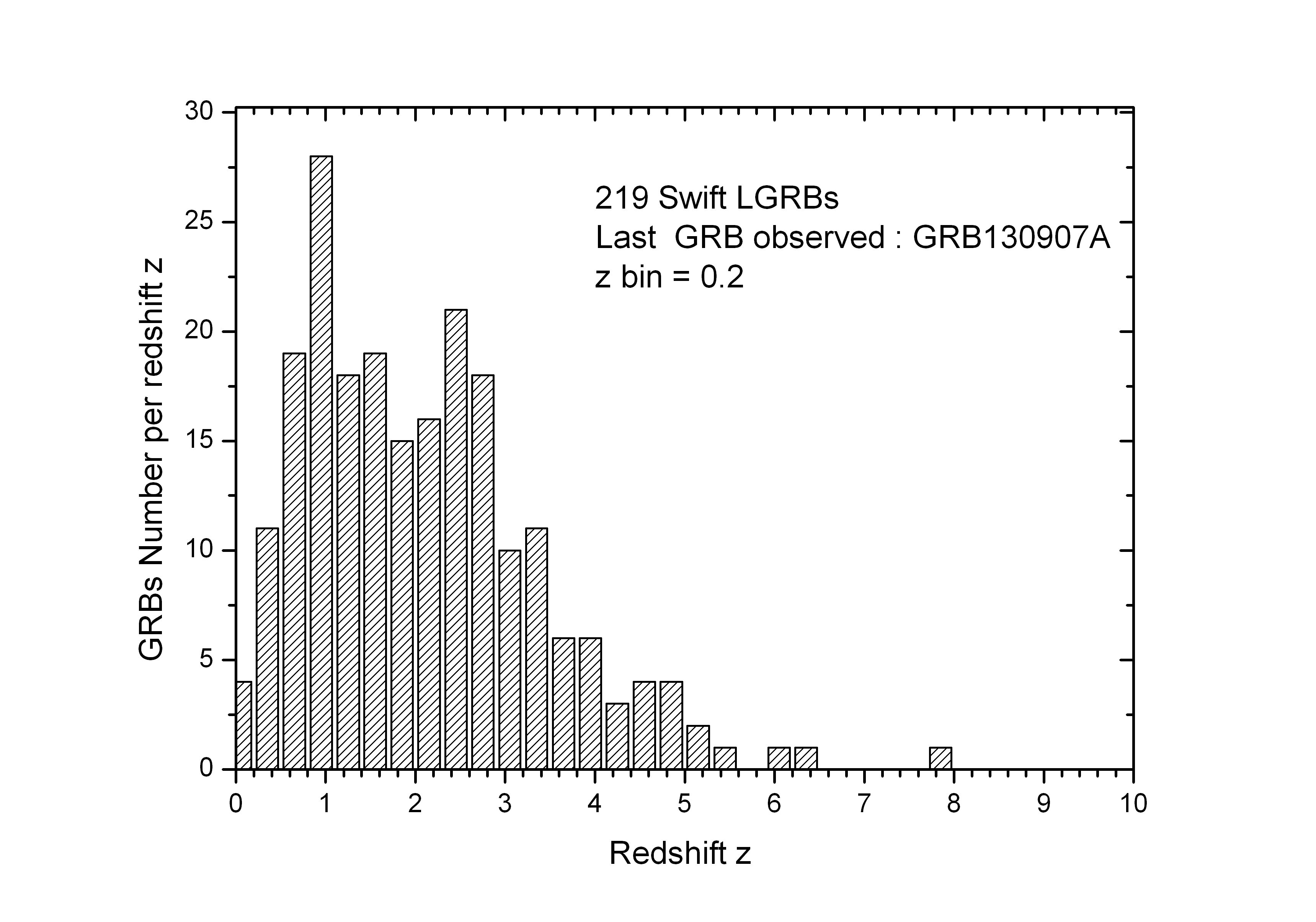}
        \caption{\emph{GRB's distribution  with redshift z, with a bin of 0.2}}
        \label{fig1}
\end{figure}

  In Figure \ref{fig1} we plot the  distribution of 219 LGRBs
  detected by \textit{Swift} up to 25/09/2013. Of these, we select the bursts
  that have an energy $E^{obs}_p$ in the \textit{Swift} range [15-150] keV, with a relative
  accuracy of at least 70\%. This selection filter leaves us with 71 GRBs. Among these, 57 have
both $L_{iso}$ and $E_{p;i}$ values and thus can be used for
testing the Amati relation, while 56 have $E_{iso}$ and $E_{p;i}$
values and thus can be used for testing the Yonetoku relation. 42
GRBs are common to both relations.  The last constraint that we
apply in selecting \textit{Swift} bursts is the photon
  flux which must be more than the threshold $P_{ph} = 2.6 ~\mathrm{ph/cm^2/s}$.
  The sample we obtain is one of 27 ``good'' GRBs for our analysis.
  We note that these 27 bursts have been very strictly selected.
  The percentage of bursts that obey the observational constraints is
  of the order of 4\% of the total number of LGRBs, and it is 13\% of LGRBs with well-determined redshift.

\section{Spectral Analysis}
The 1-sec photon flux at the peak can be found in the
\textit{Swift} data. This flux gives the total number of photons
per unit area and per unit time, regardless of their individual
energies. The \textit{Swift} energy spectrum is divided into four
bands: 15-25 keV, 25-50 keV, 50-100 keV, 100-350 keV. The spectrum
(photons per unit area, time, and energy) can be fitted by one of
three well known functions: the Band function \citep{band:92}, the
cut-off power law (CPL), and the power law (PL). The first two are
known to give very similar chi-square ($\chi^2$) values when
fitting the spectra. In the aforementioned \textit{Swift} data web
sites, only the PL and the CPL functions are given for each burst;
thus, in this work, we have chosen to use the cut-off power law,
which is characterized by two spectral parameters: the observed
peak energy, $E^{obs}_p$, and the spectral index alpha ($\alpha$).
\begin{equation}\label{eq1n0}
     N(E)=N_0(\frac{E}{E_n})^{\alpha}~~e^{-E/E_0},
   \end{equation}
where $E_n=50$ keV, a mid-range value in the interval 15-150 keV
which is only used for normalization purposes and $E_0 =
E^{obs}_p/(2+\alpha)$. The spectral parameters $E^{obs}_p$ and
$\alpha$ in the CPL function are not necessarily the same as those
of the Band law for a given burst. This observation has a direct
effect on the two correlation relations since both of them depend
on $E_{p,i}$.

To calculate the peak energy flux, expressed in
$\mathrm{erg/cm^{2}/s}$, we use the spectral parameters that
characterize the peak photons. This is given under the heading
``1-s peak spectral analysis" of the \textit{Swift} data page. We
only take what relates to the CPL spectrum. Using the CPL law, the
observed peak fluxes $P_{ph}$ that are given in Table
(\ref{tabzh3}), are calculated theoretically using the following
expression:
\begin{equation}\label{eq31zga}
P_{ph}
=N_0\int_{E_{min}}^{E_{max}}(\frac{E}{E_n})^{\alpha}~~e^{-E(2+\alpha)/E^{obs}_p}~~
dE,
\end{equation}
where $E_{min}$ = 15 keV, $E_{max}$ = 150 keV. $P_{ph}$, $\alpha$
and $E^{obs}_p$ being given in Table (\ref{tabzh3}). In
Eq.\ref{eq31zga}, the only unknown is the normalization constant
$N_0$, which we determine by numerically integrating the previous
function, i.e.
\begin{equation}\label{eqn02}
N_0 =\frac{P_{ph}}{\int_{E_{min}}^{E_{max}}(\frac{E}{E_n})^{
\alpha}~~e^{\frac{-E(2+\alpha)}{E^{obs}_p}}~ dE}.
\end{equation}
The peak energy flux, denoted by $F_{\gamma}$ and calculated in
$\mathrm{erg/cm^{2}/s}$, is calculated numerically through the
following equations:
\begin{eqnarray}\label{eq31}
F_{\gamma}&=& \int_{E_{min}}^{E_{max}}E N(E) dE,\\
& =&N_0\int_{E_{min}}^{E_{max}}
E(\frac{E}{E_n})^{\alpha}~~e^{\frac{-E(2+\alpha)}{E^{obs}_p}} dE, \label{eq31bb}\\
 & =&N_0~K~E_n~\int_{E_{min}}^{E_{max}}
(\frac{E}{E_n})^{\alpha+1}~e^{\frac{-E(2+\alpha)}{E^{obs}_p}}~ dE.
\end{eqnarray}
A factor  $K=1.6~10^{-9}$ is introduced to make the keV-erg
conversion and  $E_n=50 ~\mathrm{keV}$.

The bolometric luminosity of the 1-second isotropic peak, denoted
$L_{iso}$, is the maximum energy radiated per unit time in all
space. It is calculated by integrating the $EN_E$ function in the
energy band corresponding to the observed gamma radiation band in
the source's frame, i.e.$E_1$ = 1 keV to $E_2$ = $10^4$ keV. And
because of  cosmological effects, the corresponding observed
energy band is: $ E_1/(1+z)$ to $E_2/(1+z)$.\\ Thus, the
k-corrected $L_{iso}$ is calculated via:

\begin{eqnarray}\label{eq:LZ}
   L_{iso}&=&4~\pi~d_L^2 ~\int_{E_{1}/(1+z)}^{E_{2}/(1+z)}E N(E)dE,\\
      &=& 4~\pi~d_L^2~F_{\gamma}~k_c.
 \end{eqnarray}
 Here $L_{iso}$ is k-corrected with the method developed by
 \citep{bloom:01}. Indeed, in Eq.\eqref{eq:LZ} we  replace $N(E)$ by Eq.\eqref{eq1n0} and using
 Eq.\eqref{eq31bb} to express $N_0$, we obtain:
\begin{eqnarray}
k_c&=&\frac{{\int_{E_{1}/(1+z)}^{E_{2}/(1+z)}(\frac{E}{E_n})^{\alpha+1}~~e^{\frac{-E(2+\alpha)}{E^{obs}_{p}}}~
dE}}{{\int_{E_{min}}^{E_{max}}(\frac{E}{E_n})^{\alpha+1}~~e^{\frac{-E(2+\alpha)}{E^{obs}_{p}}}~
dE}},\nonumber\\
&=&\frac{\int_{E_{1}/(1+z)}^{E_{2}/(1+z)}E
N(E)dE}{\int_{E_{min}}^{E_{max}}E N(E) dE},
 \end{eqnarray}
where $k_c$ being the proper k-correction factor
\citep{{{yonetoku:04},rossi:08},{elliott:12}}.

These integrals are performed numerically using the time-resolved
spectral parameters given by \textit{Swift}.  The cosmological
distance $d_L$ is expressed by the following equation:
 \begin{equation}
  d_L=\frac{(1+z)c}{H_0}\int_0^z
  \frac{dz'}{\sqrt{\Omega_M(1+z')^3+\Omega_L}}.
 \end{equation}
We adopt the following cosmological parameters:\\
$\Omega_M$ = 0.27, $\Omega_L$ = 0.73, and $H_0$ = 70
$\mathrm{km/s/Mpc}$ \citep{komatsu:09}.

The total isotropic energy, denoted $E_{iso}$, which is emitted by
a gamma-ray burst in all space, is calculated using the fluences
($\mathrm{erg/cm^{2}}$) given by the detectors in the
\textit{Swift} energy band [15- 150] keV. To calculate this, we
use a cut-off power law spectrum with time-averaged spectral
parameters ($\alpha_m, E_{pm}$) obtained from the \textit{Swift}
data. Using the CPL function, the fluence $S_{obs}$, given in the
fourth column in table (\ref{tabzh3}), can be theoretically
calculated using the following equation:
\begin{eqnarray}
S_{obs} &=&\int_{E_{min}}^{E_{max}} E~N_i(E) ~~dE, \nonumber\\
&=&N'_0~E_n~T_{90}^{obs}\int_{E_{min}}^{E_{max}}(\frac{E}{E_n})^{\alpha_m+1}~~e^{\frac{-E(2+\alpha_m)}{E^{obs}_{pm}}}~~
dE, \nonumber\\
 &=&N'\int_{E_{min}}^{E_{max}}(\frac{E}{E_n})^{\alpha_m+1}~~e^{\frac{-E(2+\alpha_m)}{E^{obs}_{pm}}}~~
 dE\label{eq32zgw},
\end{eqnarray}
with $E_{min}$ = 15 keV, $E_{max}$ = 150 keV and $E_n$ = 50~keV.
$\alpha_m$ and $E^{obs}_{pm}$ are given in Table (\ref{tabzh3}),
for the time-averaged spectrum. $N_i(E)$ is the time-integrated
spectrum calculated via the product of the time-averaged spectrum
by $T_{90}^{obs}$, the observed duration of the GRB:
 \begin{eqnarray}
  N_i(E)&=&T_{90}^{obs}\times\bigg\{
  \frac{1}{T_{90}^{obs}}\int_0^{T_{90}^{obs}} N(E,t) dt\bigg\}, \nonumber\\
        &=&T_{90}^{obs}~~\bigg\{N'_0~(\frac{E}{E_n})^{\alpha_m}~~e^{\frac{-E(2+\alpha_m)}{E^{obs}_{pm}}}\bigg\}.
 \end{eqnarray}
 In
equation \ref{eq32zgw} the only unknown is the normalization
constant N'; it is determined by numerical integration, i.e.:
\begin{equation}
N'
=\frac{S_{obs}}{\int_{E_{min}}^{E_{max}}(\frac{E}{E_n})^{\alpha_m+1}~~e^{\frac{-E(2+\alpha_m)}{E^{obs}_{pm}}}~
dE}.
\end{equation}
Thus, N' is used to deduce k-corrected $E_{iso}$ :
\begin{eqnarray}
  E_{iso}&=&\frac{4\pi~d_L^2}{1+z}~\int_{E_{1}/(1+z)}^{E_{2}/(1+z)}E~N_i(E)
  dE, \nonumber\\
  &=& \frac{4~\pi~d_L^2}{1+z} S_{obs}~~k'_c.
 \end{eqnarray}

The (1 + z) factor is a cosmological correction that is needed
because one must use $T^s_{90}$ (the GRB's duration in the
source's frame) instead of $T^{obs}_{90}$ : \mbox{$T^s_{90}~=~
T^{obs}_{90}/(1 + z)$;} [$E_1$ = 1 keV ; $E_2$ = $10^4$ keV] is
the energy band in the source's frame. $k'_c$ is the k-correction
factor calculated with the parameters of the time-averaged
spectrum:

\begin{eqnarray}
  k'_c &=&\frac{{\int_{E_{1}/(1+z)}^{E_{2}/(1+z)}(\frac{E}{E_n})^{\alpha_m+1}~~e^{\frac{-E(2+\alpha_m)}{E^{obs}_{pm}}}~
dE}}{{\int_{E_{min}}^{E_{max}}(\frac{E}{E_n})^{\alpha_m+1}~~e^{\frac{-E(2+\alpha_m)}{E^{obs}_{pm}}}~
dE}},\nonumber\\
  &=&\frac{\int_{E_{1}/(1+z)}^{E_{2}/(1+z)}E N_i(E)dE}{\int_{E_{min}}^{E_{max}}E
  N_i(E)dE}.
\end{eqnarray}

\section{Results}
\subsection{Distribution of $F_{\gamma}$  $E_{iso}$ and $L_{iso}$}
\begin{figure}[t]
       \centering

         \includegraphics[angle=0, width=0.45\textwidth]{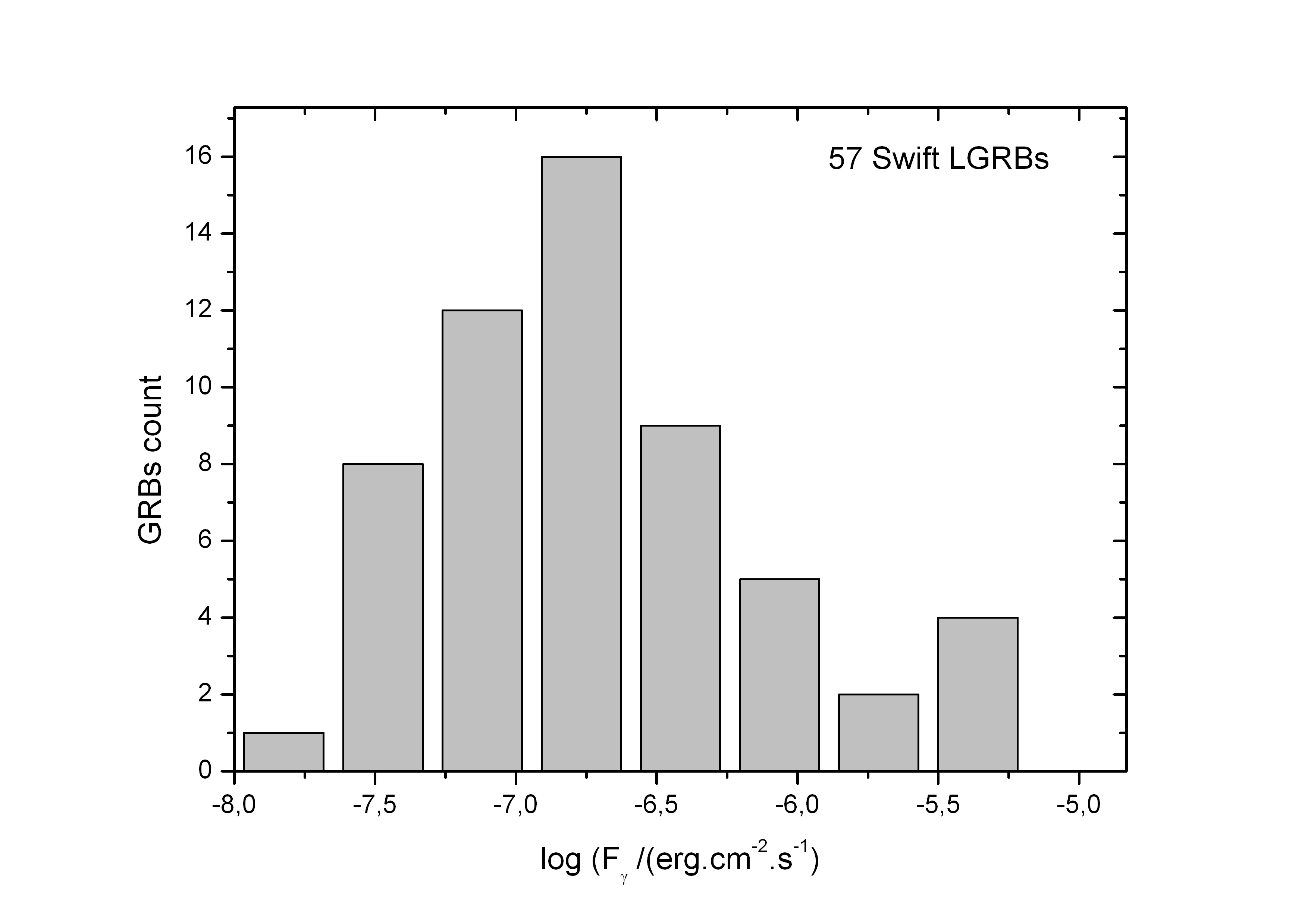}
        \caption{\emph{Histogram of LGRBs in terms of the peak energy flux $F_{\gamma}$.
        We split the interval from log($F_{\gamma}$) = 8 to log($F_{\gamma}$) = 5 into 8 bins. 57 \textit{Swift} LGRBs were used in this histogram.
        Here, we  did not consider the constraint on the threshold on the flux : $P_{ph}\geq 2.6~\mathrm{ph/cm^{2}/s}$}}
        \label{fig2}
\end{figure}
\begin{figure}[t]
       \centering
       \includegraphics[angle=0, width=0.45\textwidth]{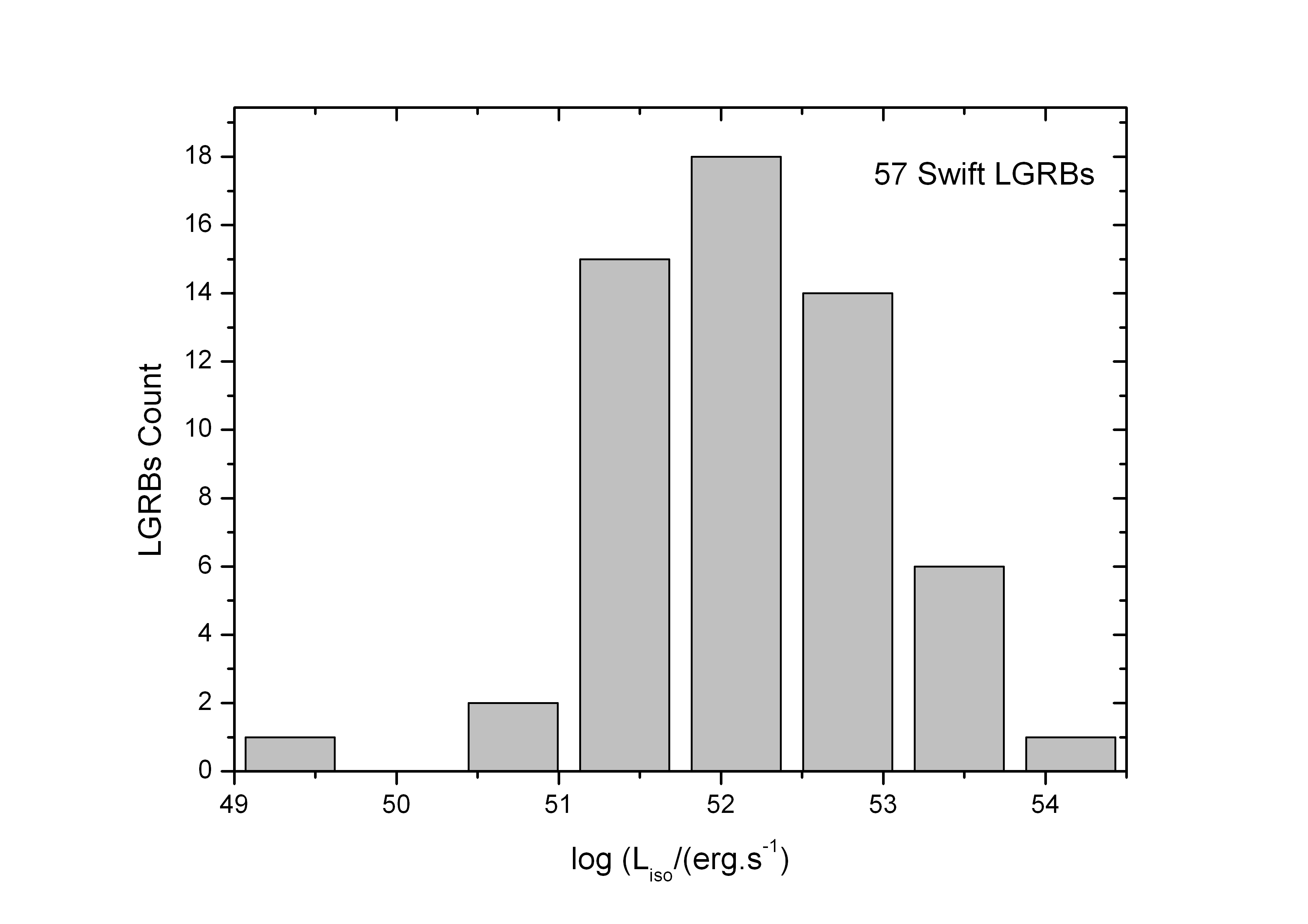}
        \caption{\emph{Histogram of LGRBs in terms of the peak  isotropic luminosity.
        We split the interval from log(Liso) = 49 to 55 into 8 bins.
        57 \textit{Swift} LGRBs were used. Here, we  did not consider the constraint: $P_{ph}\geq 2.6~\mathrm{ph/cm^{2}/s}$}}
        \label{fig3}
\end{figure}
\begin{figure}[t]
       \centering
       \includegraphics[angle=0, width=0.45\textwidth]{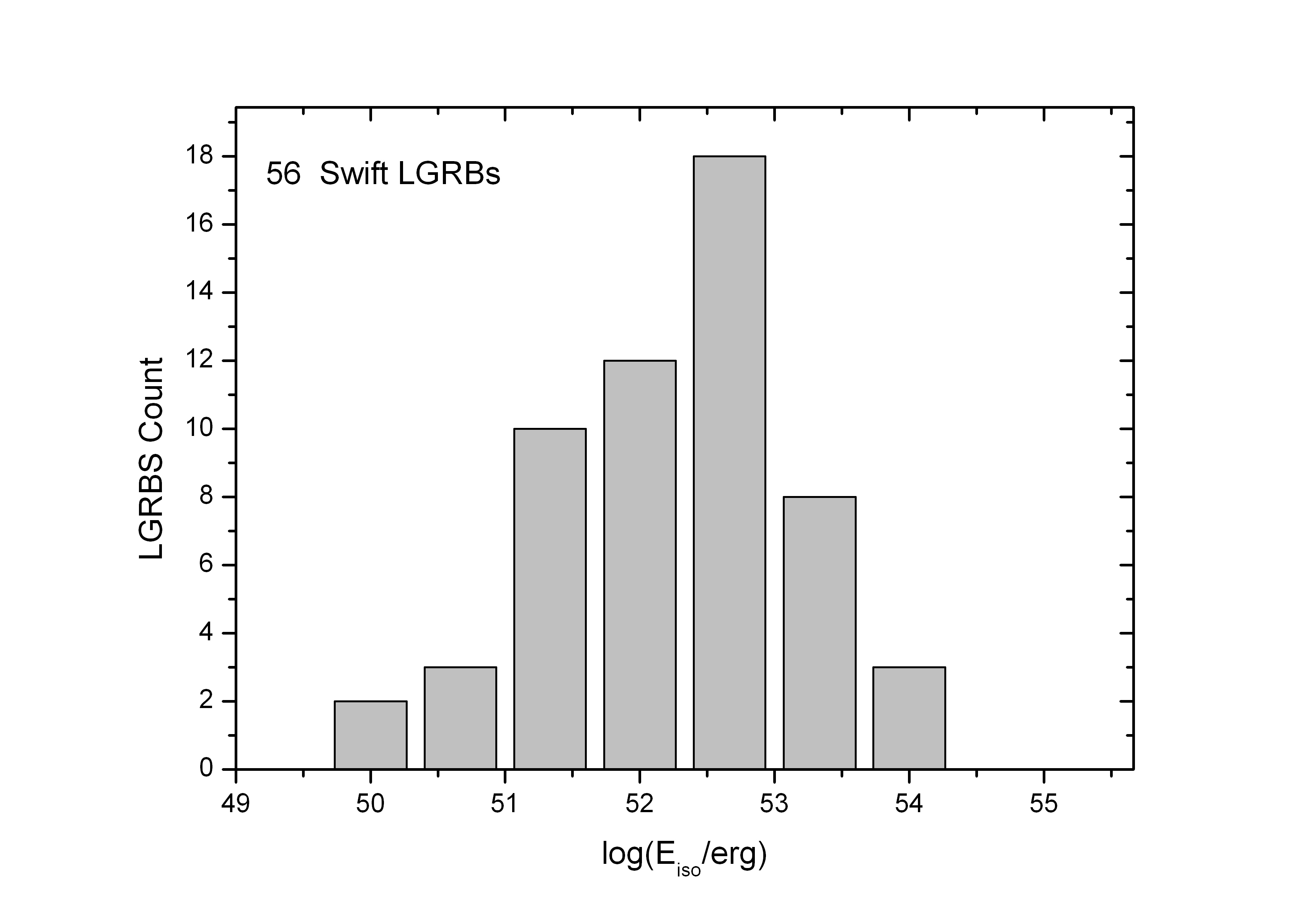}
        \caption[$E_{iso}$.]{\emph{Histogram of LGRBs in terms of the isotropic energy.
        We split the interval from $\log{E_{iso}}$ = 49 to 55 into 7 bins. 56 \textit{Swift} LGRBs
        were used. Here,  we  did not consider the constraint: $P_{ph}\geq 2.6~ph/cm^{2}/s$}}
        \label{fig4}
\end{figure}

In the previous sections we  numerically evaluated the energy flux
denoted by $F_{\gamma} \mathrm{(erg/cm^{2}/s)}$, the peak
isotropic bolometric luminosity, $L_{iso}$, and the isotropic
energy $E_{iso}$. Uncertainties over these quantities are
estimated using the Monte Carlo method. We have plotted the
distributions $P$, $E_{iso}$, $L_{iso}$ in Figures \ref{fig2},
\ref{fig3}, and \ref{fig4} respectively. We also obtained the
distributions of the two physical quantities $E_{iso}$ and
$L_{iso}$ in the sources' reference frames. In Figure \ref{fig4},
we note that $E_{iso}$, which represents the total energy released
by the burst during its entire activity, follows a lognormal
distribution, previously known \citep{preece:00}, with a mean
equal to $1.7\times10^{52}\mathrm{erg}$. Most gamma-ray bursts
that are detected by other satellites are characterized by this
average value.

This luminosity is different from the time-resolved peak
luminosity that was calculated above.

\begin{figure}
       \centering
       \includegraphics[angle=0, width=0.45\textwidth]{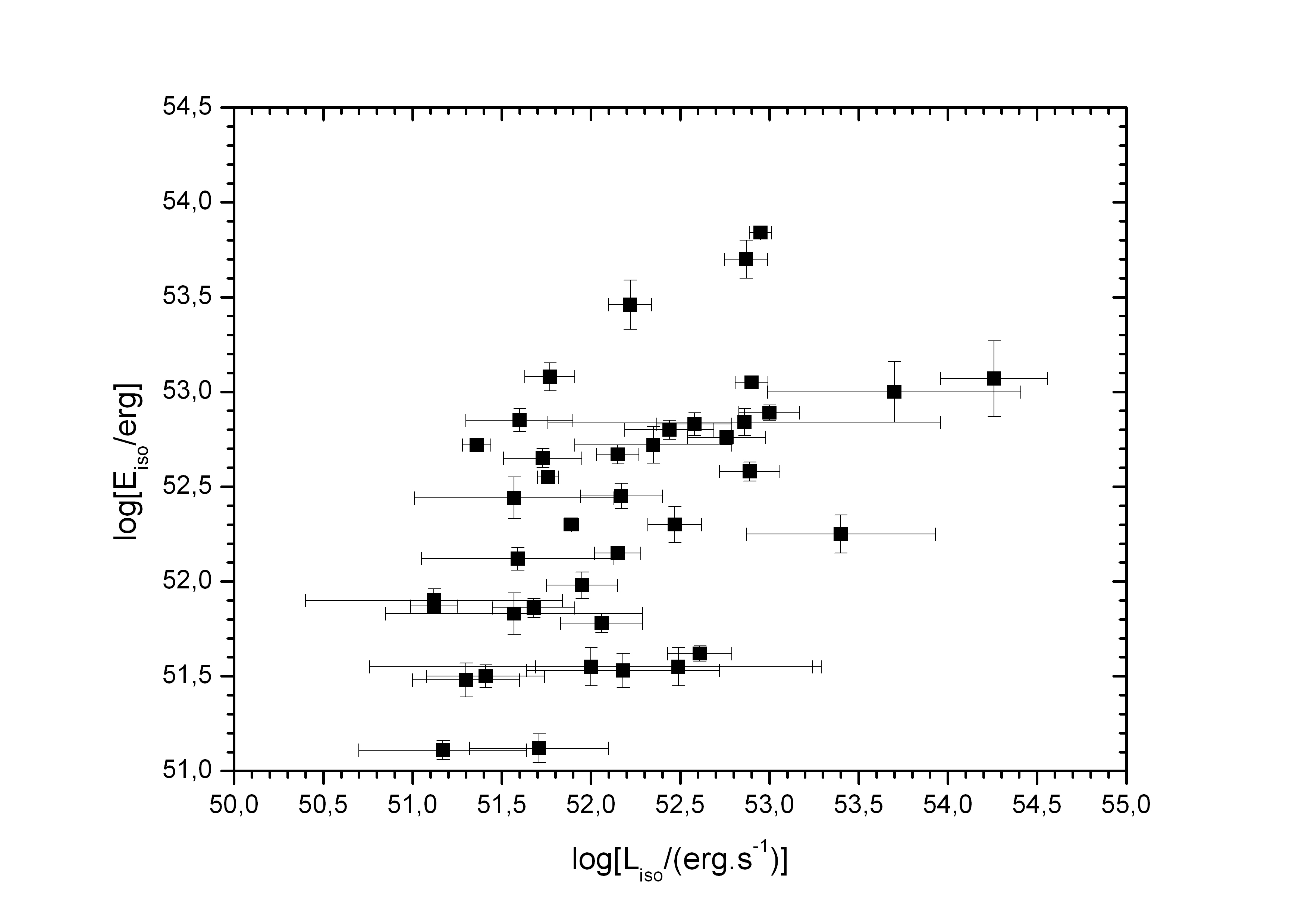}
        \caption[flux-$L_{iso}$.]{\emph{Plot of $\log{E_{iso}} $~vs.~~ $\log{L_{iso}}$:
        42 \textit{Swift} LGRBs have been used in this plot(Tab.\ref{tabzh1zga}). Here, we  did not consider the constraint: $P_{ph}\geq 2.6~ph/cm^{2}/s$. The error bars over $L_{iso}$ are much larger than those over $E_{iso}$ because
        of the size of the error bars over the peak flux.}}
        \label{fig5}
\end{figure}

In Figure(\ref{fig5}) we present the correlation between $L_{iso}$
and $E_{iso}$: a burst that releases a large amount of energy is
characterized by high luminosity. We note that these two
quantities are correlated with a wide dispersion of the
observational data.

\subsection{Correlation Relations}
One of the most debated issues regarding gamma-ray bursts (GRB )
is the existence of a correlation relation between the spectral
parameters of the prompt emission and either the total energy or
the luminosity. Three robust correlations have been identified but
not yet confirmed. Each involves the peak energy $E_p$ of the
spectrum $\nu~\textit{f}_{\nu}$ , estimated in the source's frame.
This quantity is strongly correlated with: (a) the total isotropic
energy $E_{iso} $\citep{{amati:02},{amati:06}}, (b) the isotropic
luminosity $L_{iso} $\citep{yonetoku:04}, and (c) the total energy
collimated with the opening angle  $\theta$  which is denoted by
$E_{\theta}$ \citep{ghirlanda:04}. The opening angle is inferred
from the observed break in the temporal profile of the afterglows.
In the BATSE observations, $\theta$  did not exceed ten degrees,
while \textit{Swift} observations of afterglows do not show, in
most cases, such a break. Such correlations apply only in long
GRBs. The spectral energy correlations have important implications
for both the theoretical understanding of GRBs and for
cosmological applications \citep{{{ghirlanda:04}, ghirlanda:05}}.
 \subsubsection{The $ E_{iso}-E_{p,i}$ Relation}
The relation between the energies $E_{p,i}$ and $E_{iso}$,
discovered by  \cite{amati:02} has been the subject of several
publications, even though it has not yet been fully confirmed.
This topic has had several controversies. In 2002, \cite{amati:02}
came up with this relationship from a sample of 12
\textit{Beppo-SAX} GRBs with well-determined redshifts. These
researchers showed that there is a purely empirical relation
between $E_{p,i}$, the peak energy of the photon spectrum $\nu
\textit{f}_{\nu}$ of the prompt emission, as measured in the
source's frame and the total equivalent isotropic energy $E_{iso}$
that is emitted in the energy band [1-$10^4$ keV], that is the
energy radiated by the source in this energy range, assuming an
isotropic emission. This relation requires a redshift measurement.
The redshift is necessary to know the intrinsic properties of the
source, as the intrinsic peak energy is given by the relation:
\begin{equation}
 E_{p,i} = E_p^{obs}\times(1+z),
\end{equation}
where $E^{obs}_p$ is the peak energy observed by the\\
\textit{Swift/BAT} detectors. The Amati relation is given by:
\begin{equation}\label{eq37z}
 \frac{E_{p,i}}{keV}=K\times (\frac{E_{iso}}{10^{52}~erg})^m,
\end{equation}
where K and m are constants.

For the original Amati relation, $K\approx 95$ and $m\approx0.5$.
This relation can be used to constrain cosmological parameters as
well as different models aiming to explain the prompt emission. It
can also provide information on the nature of the various
subclasses of gamma-ray bursts (e.g., LGRB, SGRB, etc.)

\begin{figure}
       \centering
       \includegraphics[angle=0, width=0.45\textwidth]{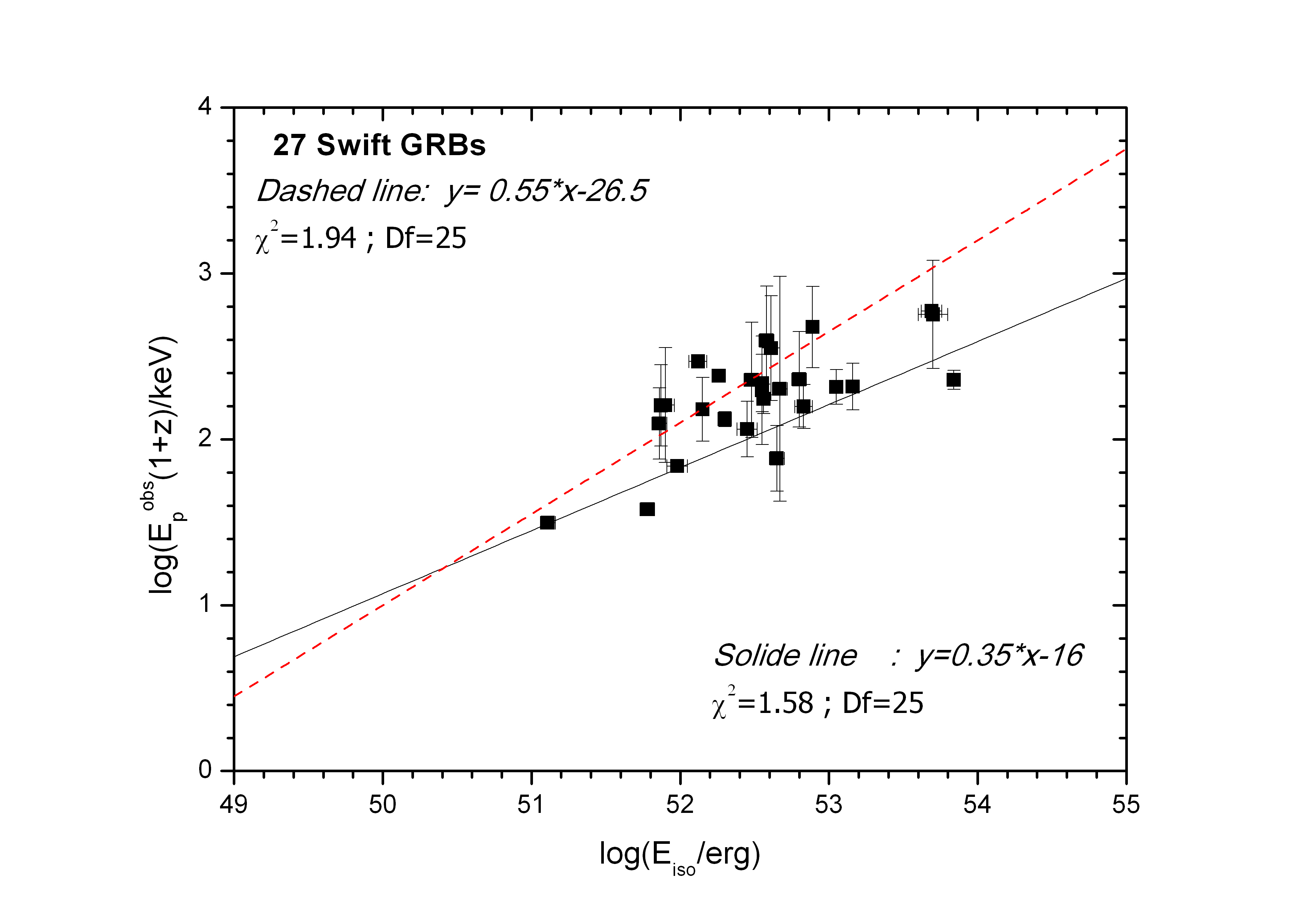}
        \caption[$E_p$-$E_{iso}$.]{\emph{The $E_{iso}-E_{p,i}$ correlation using 27 \textit{Swift} LGRBs that satisfy all the constraints that we set.}}
        \label{fig6}
\end{figure}

We plot our results for the Amati relation in Figure (\ref{fig6}).
These plots are for 27 bursts detected by \textit{Swift/BAT} with
well determined redshifts. We plot $\log{E_p}$ as a function of
$\log{E_{iso}}$. We represented two extreme lines, in fitting the
27-point distribution. From these two lines we have deduced the
mean values for the line's slope  and intercept. In Table
(\ref{EisoEp}) we give the constants K and m of  Eq.\ref{eq37z},
as obtained from the following expressions:
\begin{eqnarray}
  K&=&10^{(52a+b)},\\
  m&=&a.
\end{eqnarray}
We also give the original results of  \cite{amati:02}.
\begin{table}[ht]
  \centering
\caption{$E_{iso}-E_{p,i}$ correlation}\label{EisoEp}
  \begin{tabular}{ccccc}
    \hline
    ~  & &This work & & Amati   \\
    ~  & high & average & low & et al (2002)\\
    \hline
    K  & 126& 141 & 159& 95 \\
    m  & 0.35 & 0.45 & 0.55 & 0.5\\
    \hline
  \end{tabular}

  \end{table}

 \subsubsection{The $L_{iso}-E_{p,i}$ Relation}
 \begin{figure}
       \centering
       \includegraphics[angle=0, width=0.45\textwidth]{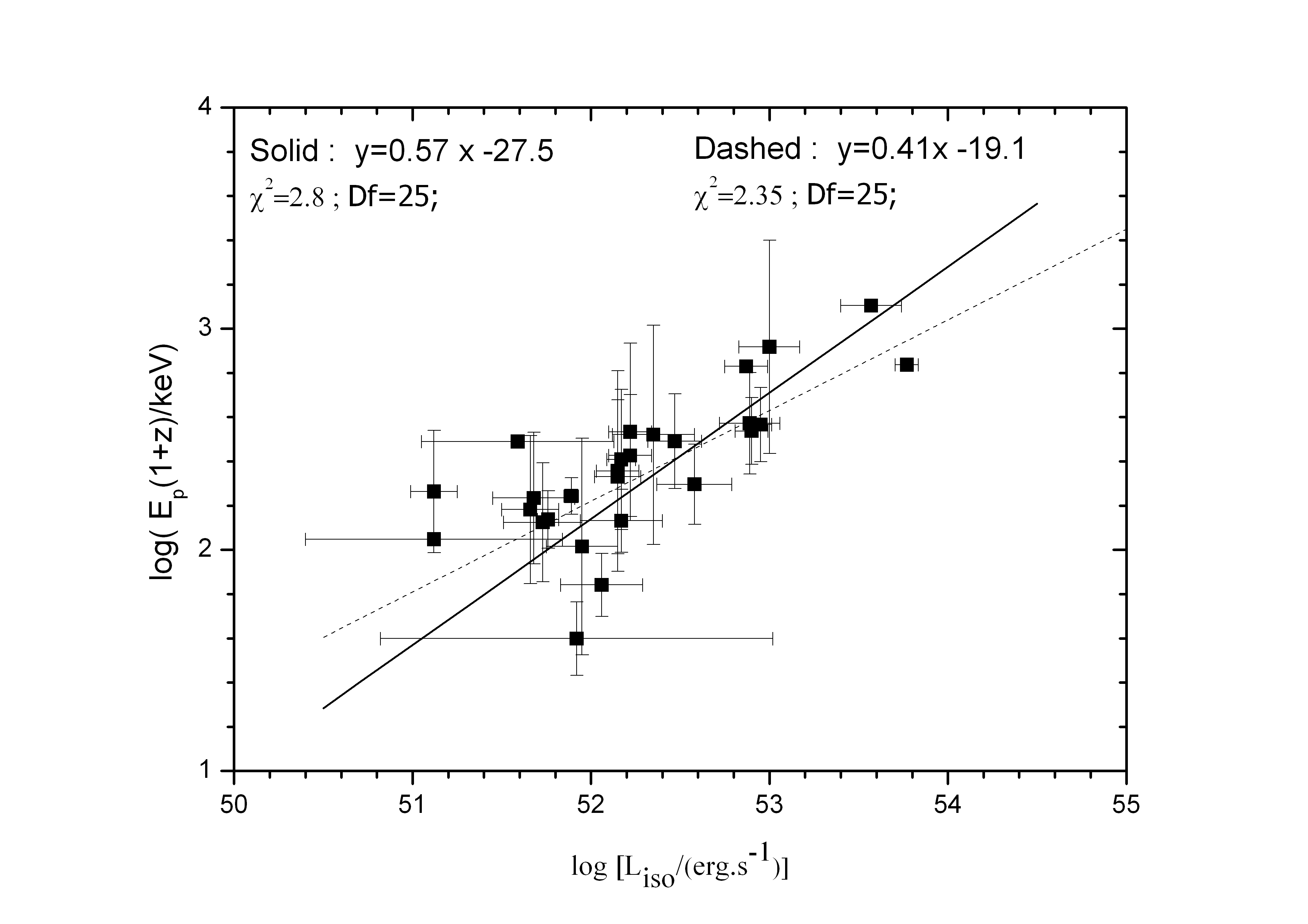}
        \caption[$E_p$-$L_{iso}$.]{\emph{The $L_{iso}-E_{p,i}$ correlation using 27 \textit{Swift} LGRBs that satisfy all the constraints that we set}}
        \label{fig7}
\end{figure}
 The relation between the energy $E_{p,i}$ and the isotropic luminosity at peak time,
 with well-determined redshifts, was found by  \cite{yonetoku:04}.
 It was expressed  as:
\begin{equation}\label{eq37}
 \frac{L_{iso}}{10^{52}~erg/s}=A~\bigg{[}\frac{E_{p,i}}{
keV}\bigg{]}^{p}.
\end{equation}
In Figure(\ref{fig7}) we plot $E_{p,i}$ vs. $L_{iso}$ in log-log
scale for 27 \textit{Swift} bursts with well determined redshifts.
We have drawn two extreme straight lines which represent
"brackets". From these two lines we deduced the mean values for
the slope and the intercept (a and b). In Table (\ref{LisoEp}) we
give the constants A and p of Eq.\ref{eq37}, which are obtained
from the following relations:
\begin{eqnarray}
  A&=&10^{-\frac{b}{a}+52},\\
  p&=&a^{-1}.
\end{eqnarray}
For comparison, we also show the original results of
\cite{yonetoku:04}.
\begin{table}[t]
  \centering
  \caption{$L_{iso}-E_{p,i}$ correlation.}\label{LisoEp}
  \begin{tabular}{ccccc}
    \hline
    ~  & &This work & & Yonetoku    \\
    ~ & high & average & low & et al (2004)\\
    \hline
    A$\times10^{-5}$  & $0.38$ & $ 3.6$ & $ 17.6$ & $4.29 \pm 0.15$\\
    p  & 2.44 & 2.04 & 1.75 & $1.94\pm 0.19$\\
    \hline
  \end{tabular}
\end{table}
\subsubsection{Evolution of ~~$L_{iso}$ and $E_{iso}$ with Redshift}
\begin{figure}[t]
         \centering
       \includegraphics[angle=0, width=0.45\textwidth]{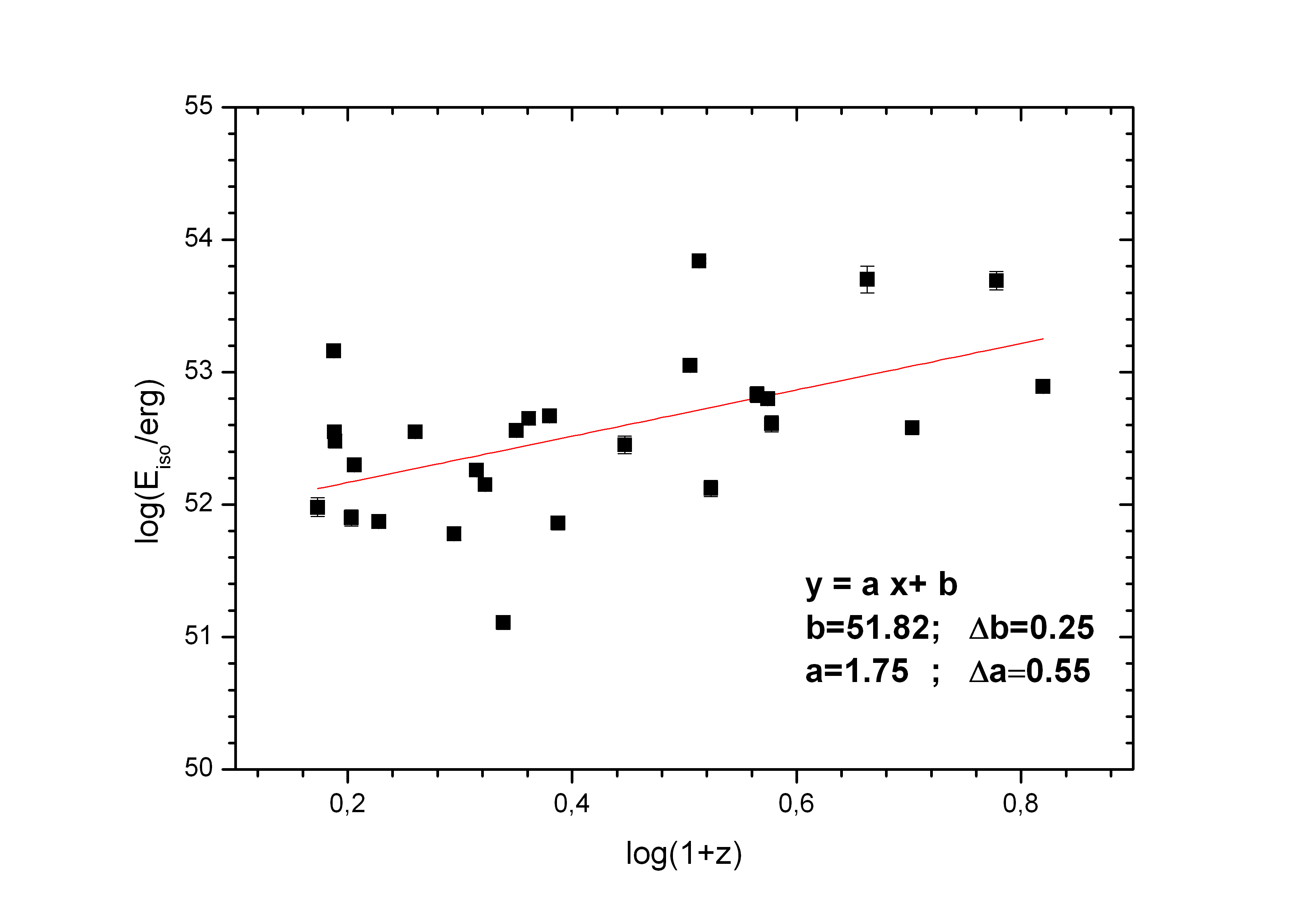}
       \hspace{0.1cm}

        \caption[$E_{iso}$ evolution with z.]{\emph{Evolution of $E_{iso}$ with the redshift z using 27 \textit{Swift}
        LGRBs that satisfy all the constraints that we set.}}
        \label{fig8}
\end{figure}
\begin{figure}[t]
         \centering
\includegraphics[angle=0, width=0.45\textwidth]{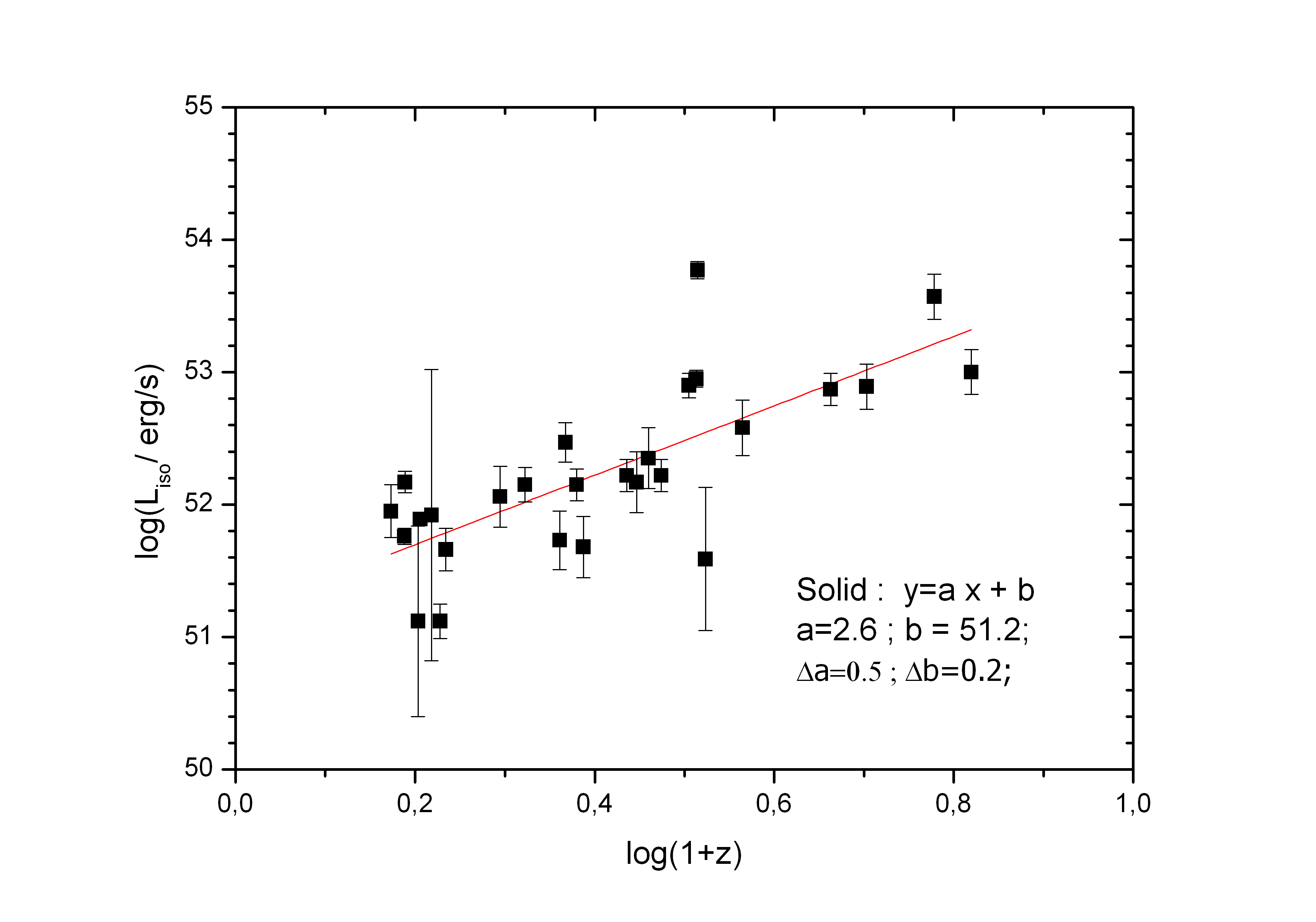}
    \caption[$L_{iso}$ evolution with z.]{\emph{Evolution of $L_{iso}$with the redshift z using 27 \textit{Swift} LGRBs
    that satisfy all the constraints that we set.}}
        \label{fig9}
\end{figure}

In the sample of 27 \textit{Swift} LGRBs, we find an interesting
evolution of the isotropic energy $E_{iso}$ in terms of the
redshift z. We plot the data in Figure (\ref{fig8}), showing a
trend between $E_{iso}$ and z, a trend which can be expressed by
the following equation:
\begin{equation}
\frac{E_{iso}}{erg}=10^{51.82\pm 0.25}~(1+z)^{1.75\pm 0.55}.
\end{equation}
This result is in good agreement with the recently published paper
\citep{salvaterra13b}. We also find a similar trend between
$L_{iso}$ and z, (Figure~\ref{fig9}), a trend which can be
expressed by the following equation:
\begin{equation}
\frac{L_{iso}}{erg/s}=10^{51.2\pm 0.2}~(1+z)^{2.6\pm 0.5}.
\end{equation}
It is indeed logical to find a (1+z) dependence in the ratio of
$L_{iso}$ and $E_{iso}$ due to the cosmological effect on the
duration  $T_{90}^{obs} = T_{90}^{s}(1+z)$, which affects only the
luminosity.

\section{Discussion}
In this section we present a brief review of recent studies that
have dealt with the Amati and Yonetoku relations in order to put
our study into proper perspective. Some studies \citep{{zhang:12},
{tsutsui:13}} have lately considered whether these correlations
apply to both short and long bursts. It had previously been
thought that the Amati relation applies only to LGRBs, whereas the
Yonetoku relation applies to both. In the study by
\citep{zhang:04} the authors used a sample of 148 LGRBs and 17
SGRBs to investigate this issue for the Yonetoku relation. The
results obtained indicate that both the LGRB and SGRB groups seem
to adhere to the correlation with the same best-fit:
$L_{iso}\propto E_{p,i}^{1.7}$. This implies that the radiation
mechanism is similar for short and long bursts, and probably has a
quasi-thermal origin in which most of the energy is dissipated
close to the central engine. On the other hand, the study by
\cite{tsutsui:13} considered both the Amati and Yonetoku relations
but for short bursts only. The authors first clarified the
sometimes ambiguous issue of when a burst is to be considered
short. They then distinguished between "secure" and "misguided"
SGRBs. Out of an initial sample of 13 bursts, 8 were found to be
"secure". With these 8 bursts they were able to obtain good fits
and to show that both the Amati and Yonetoku relations apply;
however, for a given $E_{p,i}$, $E_{iso}$ is dimmer by a factor of
about 100, and $L_{iso}$ is dimmer by a factor of about 5 than
that known for LGRBs.

Other studies have looked at the possible redshift evolution of
these correlations. The study by \cite{geng:13} used a sample of
65 bursts to investigate the possible redshift dependence of the
low-energy index, $\alpha$, in the Band function. Their results
indicate that such a dependence does exist. Although we did not
utilize the Band function in our study, since we used a CPL, our
results for the redshift dependence of $E_{iso}$ and $L_{iso}$ are
in qualitative agreement with what was found by \cite{geng:13}.

The study by \cite{nava:12} used a sample of 47 GRBs to
investigate the robustness of the Amati and Yonetoku relations,
and also to look into their possible redshift evolution. Although
the authors found some outliers, their conclusion was that these
relations are genuine and are not due to selection effects.
However, they also found no evolution of these correlations with
redshift. This final result is in agreement with a recent study
\citep{azzam:13} in which the authors investigate the possible
redshift evolution of a sample of 65 bursts by binning the data
and carrying out the proper z-correction and k-correction. The
authors obtained good fits for the binned data, but found no
evidence for redshift evolution.

Our current study is in agreement with the above investigations in
that it confirms the existence of the Amati and Yonetoku
correlations. However, we have taken a step further by
demonstrating that applying a stricter criterion for choosing the
GRB sample in the first place, actually improves the quality of
these fits, since it reduces the dispersion that is commonly seen
in these correlations. Therefore, the proper selection of the data
sample is crucial in such studies.

\section{Conclusion}

We have conducted a statistical study of a sample of
\textit{Swift} bursts. Among the 229 LGRBs with well-determined
redshifts, we selected the 71 GRBs whose observed energy
$E^{obs}_p$ is within the energy interval 15-150 keV. Among those,
57 GRBs had $L_{iso}$ and $E_{p;i}$ values and could thus be used
to test the Yonetoku relation, while 56 GRBs had $E_{iso}$ and
$E_{p;i}$ values and could thus be used for the study of the Amati
relation. These bursts  satisfy constraints on the energy
$E^{obs}_p$. The uncertainties (error bars) on the bursts'
physical quantities were estimated using a Monte Carlo method. We
present the data for the bursts, along with the error bars, in a
summary table (\ref{tabzh1zga}). For these bursts, we plotted
$E_{iso}$ against $E_{p,i}$ and $L_{iso}$ against $E_{p,i}$,
testing the Amati and Yonetoku relations on that sample. We found
the data to be tainted with significant dispersions around the
linear trends. But by adding a condition on the peak flux, we
obtained a sample of 27 LGRBs for which we got good linearities on
those two relations.

\begin{acknowledgments}

 The authors gratefully acknowledge the use of the online \textit{Swift/BAT} table compiled by
 Taka Sakamoto and  Scott D. Barthelmy. We thank the referee for constructive comments, which led us to clarify some aspects of the paper.
\end{acknowledgments}
\bibliographystyle{spr-mp-nameyear-cnd}
\bibliography{Astro_ph_zitouni_14.bib}
\appendix
{\scriptsize {
 \centering
\begin{longtable}{|l|l|l|l|l|}
\caption{Flux $F_{\gamma}$  isotropic energy $E_{iso}$ and
luminosity $L_{iso}$ calculated from the \textit{Swift} data for a
 sample of 71 GRBs that satisfy our constraints, except for the
threshold on the photon flux.}
\label{tabzh1zga}\\
  \hline
   GRB & z & Log($\frac{F_{\gamma}}{\mathrm{erg/cm^{2}/s}}$) & Log($\frac{L_{iso}}{\mathrm{erg/s}}$) & Log($\frac{E_{iso}}{\mathrm{erg}})$  \\
  \hline
\endfirsthead

\multicolumn{5}{c}%
{{\bfseries \tablename\ \thetable{} -- continued from previous
page}} \\ \hline
  GRB & z & Log($\frac{F_{\gamma}}{\mathrm{erg/cm^{2}/s}}$) & Log($\frac{L_{iso}}{\mathrm{erg/s}}$) & Log($\frac{E_{iso}}{\mathrm{erg}})$  \\ \hline
\endhead

\hline \multicolumn{5}{|r|}{{Continued on next page}} \\ \hline
\endfoot

\hline \hline
\endlastfoot
   130610A & 2.092 &  -6.85 $\pm$  0.08  &  52.40  $\pm$  0.20 &~~\\

    130514A & 3.6& -6.67 $\pm$  0.05 &  52.87  $\pm$ 0.12 &   53.70   $\pm$ 0.10\\
    130427B & 2.78  &    &      & 52.61   $\pm$ 0.06    \\
    130420A & 1.297  & -6.39   $\pm$0.03   & 51.73   $\pm$ 0.22    &52.65   $\pm$ 0.05    \\
    130215A & 0.597  & -6.73   $\pm$0.01   & 51.12   $\pm$ 0.72    &51.90    $\pm$ 0.06    \\
    121128A  &2.2    &-5.97   $\pm$0.02    &52.90    $\pm$ 0.09  & 53.05   $\pm$ 0.03    \\
    120815A & 2.358  & -6.87 $\pm$   0.04  &  51.57 $\pm$  0.72 &   51.83 $\pm$  0.11  \\
    120811C  &2.671  & -6.58  $\pm$  0.02   & 52.58 $\pm$  0.21  &  52.83 $\pm$  0.06   \\
    120724A & 1.48   & -7.56  $\pm$  0.10 & 52.49  $\pm$ 0.8& 51.55 $\pm$  0.10 \\
    120712A & 4.15   & -6.69   $\pm$ 0.06  &  53.13 $\pm$  0.58 &    \\
    120422A & 0.28   & -7.26  $\pm$  0.14  &  49.38  $\pm$ 1.20 &49.79 $\pm$  0.08  \\
    120404A & 2.876  & -7.12   $\pm$ 0.05  &  52.18  $\pm$ 0.08 &      \\
    120326A & 1.798   &-6.55   $\pm$ 0.017 &  52.17  $\pm$ 0.23 &   52.45  $\pm$ 0.07 \\
    120118B & 2.943  & -6.86   $\pm$ 0.037 &  52.35  $\pm$ 0.44 &   52.72 $\pm$  0.10  \\
    111229A & 1.381  & -7.15  $\pm$  0.08  &  51.30  $\pm$  0.30 &51.48   $\pm$ 0.09  \\
    111228A  &0.7141&  -6.1  $\pm$   0.02 &   51.66  $\pm$ 0.16   & \\
    111107A & 2.893&     &    &   53.48 $\pm$  0.67  \\
    111008A & 5   &    &     &  53.69  $\pm$ 0.07  \\
    110808A & 1.348  & -7.53 $\pm$  0.12 &   51.41 $\pm$  0.33 &   51.50  $\pm$  0.06  \\
    110801A & 1.858  & -7.16  $\pm$ 0.05 &   51.60   $\pm$ 0.30& 52.85  $\pm$ 0.06    \\
    110715A & 0.82   &  &     &   52.55 $\pm$  0.02   \\
    110503A & 1.613  & -6.91  $\pm$ 0.03  &  51.77  $\pm$ 0.14 &   53.08  $\pm$ 0.07   \\
    110205A &1.98   &  -6.57 $\pm$  0.03 &   52.22$\pm$   0.12  & \\
    110128A & 2.339  & -7.17 $\pm$  0.1 &  51.59  $\pm$ 0.54 &   52.12 $\pm$0.06\\
    100906A & 1.727 &  -6.57 $\pm$  0.07 &   52.22 $\pm$  0.12 &  \\
    100621A & 0.542 &  -6.02 $\pm$  0.01 &  51.76  $\pm$ 0.06 &   52.55 $\pm$  0.03 \\
    100615A & 1.398 &  -6.41 $\pm$  0.03 &  52.15 $\pm$  0.12 &   52.67  $\pm$ 0.05  \\
    100513A & 4.772 &    &    &   53.00 $\pm$ 0.03  \\
    100425A & 1.755 &  -7.17  $\pm$ 0.05 &  52.00 $\pm$ 1.24   & 51.55 $\pm$  0.10 \\
    100418A & 0.624 &  &   &50.83  $\pm$ 0.05 \\
    100316B & 1.18  &  -7.12 $\pm$  0.04 &  51.17 $\pm$  0.47 &   51.11 $\pm$  0.05 \\
    091208B & 1.063 &     &   &   52.26 $\pm$  0.04  \\
    091127 & 0.49  &  -5.51  $\pm$ 0.03  &  51.95 $\pm$  0.20 & 51.98  $\pm$ 0.07 \\
    091029 & 2.752 &  -6.94 $\pm$  0.03  &  52.44 $\pm$  0.25 &   52.80 $\pm$   0.05 \\
    091018 & 0.971 &  -6.23 $\pm$  0.17  &  52.06  $\pm$ 0.23 &   51.78  $\pm$ 0.05 \\
    090927 & 1.37  &  -6.88 $\pm$  0.06  &  51.71 $\pm$  0.40 &   51.12  $\pm$ 0.08 \\
    090926B & 1.24  &     &    & 52.56  $\pm$ 0.02  \\
    090726 & 2.71  &  -7.41  $\pm$ 0.20 &    53.40  $\pm$  0.53&    52.25 $\pm$  0.10 \\

    090618 & 0.54  &    &   &  53.16 $\pm$  0.02  \\
    090424  &0.544  &    &   &   52.48 $\pm$  0.05 \\
    090423 & 8.0 &  -6.93  $\pm$ 0.04 &   52.76 $\pm$  0.22 &   52.76 $\pm$  0.04 \\

    081221 & 2.26 &   -5.83  $\pm$ 0.02 &  52.95 $\pm$  0.06&   53.84 $\pm$  0.02 \\
    081121 & 2.512&   -6.49  $\pm$ 0.16  &  54.26 $\pm$  0.30 & \\

    081118 & 2.58 &   -7.40  $\pm$  0.10 &51.57 $\pm$  0.56   &  \\
    080916A & 0.689 &  -6.61  $\pm$ 0.04 &   51.12  $\pm$ 0.13  &  51.87 $\pm$  0.03\\

    080520 & 1.545 &  -7.60  $\pm$  0.12 &   51.27  $\pm$ 1.43 &  \\
    080413B & 1.1& -5.85  $\pm$ 0.03&   52.15 $\pm$  0.13 &   52.15 $\pm$  0.03  \\
    80330   &  1.51&&&    51.64 $\pm$  0.06\\

    080310 & 2.427&   -7.11 $\pm$  0.14 &   52.49 $\pm$  0.70 &\\

    071122 & 1.14 &   -7.64  $\pm$ 0.18  &  50.33 $\pm$  0.44  &  51.46 $\pm$  0.08 \\
    071117 & 1.331 &  -5.99  $\pm$ 0.02 &  52.47 $\pm$  0.15   &  \\
    071010A & 0.98  &  -7.47  $\pm$ 0.13  &  49.90  $\pm$  0.59   & 50.70  $\pm$  0.05 \\
    070611  &2.04  &  -7.30  $\pm$  0.08  &  52.61 $\pm$  0.18   & 51.62 $\pm$  0.04 \\
    070506 & 2.31 &   -7.23   $\pm$ 0.12 &   52.18  $\pm$ 0.54   &  \\
    070419A & 0.97 &   -8.04   $\pm$ 0.18 &   49.51  $\pm$ 0.55    &51.21  $\pm$ 0.04  \\
    070129  &2.338 &  -7.48  $\pm$ 0.06 &  51.36 $\pm$  0.08   & 52.72  $\pm$ 0.03  \\

    061222B & 3.355 &  -7.08  $\pm$ 0.11  &  52.86 $\pm$  1.10 &52.84  $\pm$ 0.07   \\
    060927 & 5.6 &-6.62   $\pm$ 0.04    &53.00 $\pm$ 0.17   & 52.89 $\pm$  0.04  \\
    060926 & 3.208 &  -7.26  $\pm$ 0.11 &   52.64 $\pm$  0.70 &\\

    060908 & 1.884 &  -6.56  $\pm$ 0.05 &   52.35 $\pm$  0.23 &    \\
    060604 & 2.136 &    &    &   51.82 $\pm$  0.07  \\
    060526 & 3.21  &  -6.88  $\pm$ 0.08&    53.70 $\pm$   0.71 &   53.00 $\pm$ 0.16  \\

    060512 & 0.443 &  -7.42  $\pm$ 0.09 &  50.44  $\pm$ 1.50& 50.10  $\pm$  0.14 \\
    060510B & 4.9 &  &     & 53.48 $\pm$  0.04 \\
    060218 & 0.033 &  -7.90  $\pm$ 0.14 &  46.53 $\pm$ 1.42& 48.52  $\pm$  0.08 \\

    060206 & 4.045 &  -6.70   $\pm$ 0.03 &   52.89 $\pm$  0.17  &  52.58 $\pm$  0.05  \\
    060115  &3.53 &     &    &  52.75 $\pm$  0.06  \\
    050525A  &0.606&   -5.46  $\pm$ 0.12&    51.89 $\pm$  0.04   & 52.30  $\pm$  0.01  \\
    050416A & 0.653 &  -6.65  $\pm$ 0.10 &51.92 $\pm$  1.10 & \\
    050406  &2.44 &   -7.77  $\pm$ 0.12 &   50.93  $\pm$ 1.60 &50.90  $\pm$  0.09  \\
    050318  &1.44  &  -6.66  $\pm$ 0.03 &   51.68 $\pm$  0.23  &  51.86 $\pm$  0.05 \\
 \hline
\end{longtable}
} }
 \scriptsize {
\begin{longtable}{|l|l|l|l|l|l|l|l|l|}
\caption{Data for 71 \textit{Swift} GRBs.} \label{tabzh3}\\
  \hline
  GRB & z & $T_{90}$ &  Fluence$*10^{-7}$&$1s-flux$  & ~~~~$\alpha$&$E_p^{obs}$&~~~~$\alpha_m$&$E_{pm}^{obs}$ \\
   &   & sec &  $\mathrm{erg/cm^{2}}$&$\mathrm{ph/cm^{2}/s}$  & &keV& &keV \\
    &   &   &  &  & 1s Pic spec &1s Pic Spec&Aver.Spec& Aver.Spec \\
\hline
\endfirsthead

\multicolumn{9}{c}%
{{\bfseries \tablename\ \thetable{} -- continued from previous page}} \\
\hline  GRB & z & $T_{90}$ &  Fluence$*10^{-7}$&$1s-flux$  & ~~~~$\alpha$&$E_p^{obs}$&~~~~$\alpha_m$&$E_{pm}^{obs}$ \\
   &   & sec &  $\mathrm{erg/cm^{2}}$&$\mathrm{ph/cm^{2}/s}$  & &keV& &keV \\
    &   &   &  &  & 1s Pic spec &1s Pic Spec&Aver.Spec& Aver.Spec \\ \hline
\endhead

\hline \multicolumn{9}{|r|}{{Continued on next page}} \\ \hline
\endfoot

\hline \hline
\endlastfoot
  \hline
  130610A &  2.092 &  46.4 &   25   $\pm$   1 &      1.7 $\pm$    0.2 &    0.202  $\pm$  0.798  &    93.7  $\pm$ 65.3   &  1.026   $\pm$ 0.284   &  217.1 $\pm$  62.8 \\
  130514A & 3.6& 204 &    91   $\pm$   2   &    2.8   $\pm$  0.3  &   1.173  $\pm$  0.506  &   146.9  $\pm$  1.0 &    1.646  $\pm$  0.193 &    122.9  $\pm$ 40.2  \\
     130427B & 2.78  &  27  &    15  $\pm$    1   &    3   $\pm$    0.4  &   1.182  $\pm$  0.646  &   176.3  $\pm$  1.0 &    1.229 $\pm$   0.594  &    93.8 $\pm$  29.6     \\
    130420A & 1.297  & 123.5  &  71  $\pm$    3  &     3.4  $\pm$   0.2 &    1.030  $\pm$  0.486  &    58.0  $\pm$ 15.6  &   1.518  $\pm$  0.251  &    33.4 $\pm$   6.6 \\
     130215A & 0.597  & 65.7 &   54   $\pm$   5   &    2.5  $\pm$   0.7  &   0.162  $\pm$  2.978  &    69.9  $\pm$  1.0  &   1.126  $\pm$  0.545  &   100.5  $\pm$ 34.8 \\
    121128A  &2.2    &23.3 &   69   $\pm$   4    &   12.9  $\pm$  0.4  &   0.494   $\pm$ 0.225   &  107.7  $\pm$ 16.2   &  1.320  $\pm$  0.183  &    64.6  $\pm$  6.8 \\
    120815A & 2.358  & 9.7 &    4.9  $\pm$   0.7  &   2.2  $\pm$   0.3  &   1.222  $\pm$  1.091  &    45.7 $\pm$   1.0   &  1.033  $\pm$  1.259  &    28.6  $\pm$  1.0\\
    120811C  &2.671  & 26.8 &   30  $\pm$    3  &     4.1  $\pm$   0.2   &  1.015 $\pm$   0.458   &   53.9  $\pm$  9.8  &   1.401  $\pm$  0.303  &    42.9  $\pm$  5.7\\
    120724A & 1.48   & 72.8 &   6.8  $\pm$   1.1 &    0.6  $\pm$   0.2   & -2.910  $\pm$  0.100  &    41.1 $\pm$  10.0  &  0.534  $\pm$  1.529 &     27.6  $\pm$  7.5  \\
    120712A & 4.15   & 14.7 &   18  $\pm$    1   &    2.4   $\pm$  0.2  &   0.133  $\pm$  0.742   &   99.7 $\pm$  55.8   &  0.984 $\pm$   0.306  &   143.2 $\pm$ 158.6 \\
    120422A & 0.28   & 5.35 &   2.3  $\pm$   0.4  &   0.6   $\pm$  0.2  &   -0.558  $\pm$  2.948  &   103.6 $\pm$  34.1  &  0.398  $\pm$  1.099  &    91.5  $\pm$ 30.7\\
    120404A & 2.876  & 38.7  &  16  $\pm$    1   &    1.2  $\pm$   0.2  &   2.052 $\pm$   0.100  &    21.5  $\pm$  1.0   &  1.821  $\pm$  0.100   &  269.3  $\pm$  1.0\\
    120326A & 1.798  &69.6 &   26  $\pm$    3    &   4.6   $\pm$  0.2   &  1.127  $\pm$  0.410   &   48.4   $\pm$ 6.9   &  1.409  $\pm$  0.338   &   41.1  $\pm$  6.9\\
    120118B & 2.943  & 23.26 &  18   $\pm$   1   &    2.2   $\pm$  0.3  &   1.434  $\pm$  1.085  &    50.7  $\pm$  1.0  &   1.599  $\pm$  0.506  &    39.0  $\pm$  1.0 \\
    111229A & 1.3805  & 25.4 &   3.4  $\pm$   0.7 &    1   $\pm$    0.2 &    1.379  $\pm$  1.046 &    108.1 $\pm$   1.0  &   1.764 $\pm$   0.904  &   102.7  $\pm$  1.0\\
    111228A  &0.7141&  101.2 &  85  $\pm$    2  &     12.4  $\pm$  0.5  &   1.650 $\pm$   0.272  &    88.7 $\pm$  29.7   &  1.989  $\pm$  0.100    &   1.9  $\pm$  7.2 \\
    111107A & 2.893&   26.6 &   8.8  $\pm$   0.8  &   1.2  $\pm$   0.2  &   1.034 $\pm$   0.630  &   782.8  $\pm$  1.0   &  2.285  $\pm$  0.198   &  102.1  $\pm$  1.0 \\
    111008A & 5   &63.46 &  53   $\pm$   3  &     6.4  $\pm$   0.7 &    1.011 $\pm$   0.539 &    212.3  $\pm$  1.0  &   1.725  $\pm$  0.338  &    98.7 $\pm$   1.0 \\
    110808A & 1.348  & 48  &    3.3  $\pm$   0.8  &   0.4  $\pm$   0.2  &   0.168  $\pm$  1.113  &    65.8   $\pm$ 1.0  &   1.854 $\pm$   0.100   &   11.5 $\pm$   1.0 \\
    110801A & 1.858  & 385 &    47  $\pm$    3  &     1.1  $\pm$   0.2  &   1.646  $\pm$  0.939  &    62.5  $\pm$  1.0   &  1.615  $\pm$  0.296   &   78.6 $\pm$  26.2 \\
    110715A & 0.82   & 13  &    118  $\pm$   2  &     53.9 $\pm$   1.1   &  0.985  $\pm$  0.131  &   152.0  $\pm$ 32.6  &   1.254  $\pm$  0.116  &   119.8 $\pm$  20.8\\
    110503A & 1.613  & 10  &    100  $\pm$   4  &     1.35  $\pm$  0.06  &  0.100  $\pm$  0.356  &   111.8 $\pm$  21.9   &  0.881 $\pm$   0.254  &   133.1  $\pm$ 54.5 \\
    110205A & 1.98  & 257  &    170             &     3.6    $\pm$  0.2  & 1.219  $\pm$  0.344   &  144.6 $\pm$   -46.0  &  1.527  $\pm$  0.185  &    94.2  $\pm$   61.5\\
    110128A & 2.339  & 30.7 &7.2  $\pm$   1.4   &  0.8   $\pm$ 0.2 & 0.052  $\pm$  2.242  &    92.5   $\pm$   0.0&0.265  $\pm$  2.001  &    88.4  $\pm$  0.0\\
    100906A & 1.727 &  114.4 &  120  $\pm$   0.1 &     10.1 $\pm$   0.4   &  0.876 $\pm$   0.294 &     97.9  $\pm$ 26.9 &    1.655  $\pm$  0.149  &   108.6 $\pm$ 109.4\\
    100621A & 0.542 & 63.6  &  210  $\pm$   0.1  &   12.8  $\pm$  0.3   &  0.918  $\pm$  0.143   &   89.0  $\pm$ 11.6   &  1.814  $\pm$  0.114  &   128.3  $\pm$ 42.2\\
    100615A & 1.398 &  39   &   50   $\pm$   1   &    5.4  $\pm$   0.2  &   1.157  $\pm$  0.245  &    94.8  $\pm$ 43.5  &   1.647  $\pm$  0.176  &    84.0  $\pm$ 57.4\\
    100513A & 4.772 &  84   &   14    $\pm$  1   &    0.6  $\pm$   0.1  &   1.253  $\pm$  0.100  &   939.1 $\pm$   1.0  &   1.364  $\pm$  0.437   &  115.6  $\pm$  1.0 \\
    100425A & 1.755 & 37    &  4.7   $\pm$  0.9  &   1.4  $\pm$   0.2  &  -0.351  $\pm$  2.407  &    30.2  $\pm$  5.9  &   0.847  $\pm$  1.670   &   26.6   $\pm$ 1.0\\
    100418A & 0.6235 & 7    &   3.4  $\pm$   0.5  &   1    $\pm$   0.2 &    1.982  $\pm$  0.100  &     1.1 $\pm$   1.0 &    1.920 $\pm$   0.100  &    18.5  $\pm$  1.0 \\
    100316B & 1.18  & 3.8 &    2   $\pm$     0.2 &    1.3  $\pm$    0.1  &   1.530  $\pm$   0.730  &    30.2  $\pm$  21.7  &   1.858  $\pm$   0.496  &    14.4  $\pm$   1.0\\
    091208B & 1.063 &  14.9 &   33  $\pm$    2   &    15.2  $\pm$  1.0   &  1.315  $\pm$  0.351 &    255.5  $\pm$  1.0  &   1.595  $\pm$  0.338  &   116.9  $\pm$  1.0\\
    091127 & 0.49  &  7.1  &   90   $\pm$   3   &    46.5  $\pm$  2.7   &  1.329  $\pm$  0.399  &    69.5 $\pm$  34.2   &  1.797  $\pm$  0.280   &   46.3  $\pm$  1.0 \\
    091029 & 2.752 & 39.2 &   24  $\pm$    1   &    1.8   $\pm$  0.1   &  0.923  $\pm$  0.652  &    52.8  $\pm$ 17.7   &  1.465  $\pm$  0.275   &   61.3 $\pm$  17.6\\
    091018 & 0.971 &  4.4  &   14  $\pm$    1  &     10.3  $\pm$  0.4  &   1.296  $\pm$  0.306 &     35.2 $\pm$   5.0   &  1.765  $\pm$  0.242   &   19.2  $\pm$  1.0 \\
    090927 & 1.37  &  2.2   &  2   $\pm$    0.3 &    2  $\pm$     0.2   &  0.682  $\pm$  0.977 &     56.0  $\pm$ 36.4   &  1.303  $\pm$  0.649   &   61.8  $\pm$  1.0\\
    090926B & 1.24  &  109.7 &   73  $\pm$    2 &      3.2  $\pm$   0.3 &    0.819  $\pm$  0.594 &    177.2  $\pm$  1.0 &    0.517 $\pm$   0.236  &    78.2 $\pm$   7.0\\
    090726 & 2.71  &  67  &    8.6  $\pm$   1   &    0.7  $\pm$   0.2  &  -1.928  $\pm$  0.278  &    40.3 $\pm$   9.0  &   1.345  $\pm$  0.858   &   27.3  $\pm$  1.0\\
    090618 & 0.54  &  113.2&   1050 $\pm$   10  &    38.9 $\pm$   0.8 &    1.153 $\pm$   0.144  &   170.4  $\pm$ 68.7 &    1.414 $\pm$   0.081  &   134.7 $\pm$  19.1 \\
    090424  &0.544  & 48  &    210  $\pm$   1   &    71   $\pm$   2   &    0.915  $\pm$  0.137  &   166.0  $\pm$ 52.6 &    1.244  $\pm$  0.140  &   147.8  $\pm$ 51.3\\
    090423 & 8.0 &  10.3  &  5.9   $\pm$  0.4   &  1.7   $\pm$  0.2  &   1.199  $\pm$  0.515    &  84.5 $\pm$  28.6   &  0.765  $\pm$  0.470  &    53.2  $\pm$  7.0\\
    081221 & 2.26 &   34   &   181  $\pm$   3   &    18.2 $\pm$   0.5   &  0.749  $\pm$  0.190  &   112.9  $\pm$ 19.3   &  1.211 $\pm$   0.128   &   69.9 $\pm$   3.9\\
    081121 & 2.512&  14   &   41   $\pm$   3    &   4.4   $\pm$  1    &  -2.441  $\pm$  0.100   &   56.8  $\pm$  8.0  &   0.563  $\pm$  0.500   &  140.3 $\pm$ 126.8 \\
    081118 & 2.58 &   67  &    12  $\pm$    1   &    0.6  $\pm$   0.2 &    1.246 $\pm$   1.723  &    72.3  $\pm$  1.0 &    1.569  $\pm$  0.619   &   34.3 $\pm$  24.9\\
    080916A & 0.689 &  60  &    40  $\pm$    1  &     2.7  $\pm$   0.2 &    0.031 $\pm$   0.468 &    108.7 $\pm$  30.3 &    1.167  $\pm$  0.206  &    94.7 $\pm$  23.2\\
    080520 & 1.545 &  2.8  &   0.55  $\pm$  0.17 &   0.5   $\pm$  0.1 &    0.418 $\pm$   2.495  &    28.7  $\pm$ 21.0   &  1.742  $\pm$  1.450  &     7.1 $\pm$  13.1 \\
    080413B & 1.1& 8  &     32  $\pm$    1   &    18.7  $\pm$  0.8   &  1.005  $\pm$  0.291   &  102.0  $\pm$ 35.5   &  1.222  $\pm$  0.276   &   72.2  $\pm$ 13.9 \\
    080330 & 1.51 &61 &     3.4  $\pm$   0.8 &    0.9  $\pm$   0.2   &  1.984  $\pm$  0.372   &    1.4  $\pm$  1.0  &   1.863  $\pm$  0.100  &     7.5  $\pm$  1.0 \\
    080310 &2.4266 & 365 &    23  $\pm$    2  &     1.3  $\pm$   0.2 &    0.669  $\pm$  1.454 &     41.9  $\pm$ 17.5 &    1.945 $\pm$   0.522  &     5.5  $\pm$ 14.5\\
    071122 & 1.14 &  68.7 &   5.8  $\pm$   1.1&     0.4  $\pm$   0.2 &    1.985 $\pm$   0.100  &     3.2  $\pm$  1.0 &    1.423 $\pm$   0.937  &    79.4 $\pm$   1.0 \\
    071117 & 1.331 & 6.6  &   24  $\pm$    1   &    11.3  $\pm$  0.4  &   0.339 $\pm$   0.256  &   133.0 $\pm$  28.4  &   1.232  $\pm$  0.244  &   127.2 $\pm$  94.0\\
    071010A & 0.98  & 6   &    2   $\pm$    0.4 &    0.8  $\pm$   0.3  &   1.992  $\pm$  0.100  &     1.0  $\pm$  1.0  &   0.687  $\pm$  0.100  &    35.6 $\pm$   1.0\\
    070611  &2.04  &  12.2 &   3.91  $\pm$  0.57 &   0.82  $\pm$  0.21 &  -1.819  $\pm$  0.100  &    46.2  $\pm$  8.9   &  0.374  $\pm$  0.100  &    57.7  $\pm$  1.0\\
    070506 & 2.31 &   4.3 &    2.1 $\pm$    0.23  &  0.96  $\pm$  0.13 &  -0.045  $\pm$  1.227  &    47.3 $\pm$  12.0  &   0.924  $\pm$  0.770  &    57.8 $\pm$  50.6\\
    070419A & 0.97 &  115.6 &  5.58  $\pm$  0.83  &  0.2  $\pm$   0.1  &   1.975   $\pm$ 0.100   &    1.4  $\pm$  1.0  &   1.101  $\pm$  0.100 &     24.3  $\pm$  1.0\\
    070129  &2.338 & 460.6  & 29.8  $\pm$  2.67  &  0.55  $\pm$  0.12  &  1.445  $\pm$  0.100  &    43.5   $\pm$ 1.0   &  1.390  $\pm$  0.100  &    40.9  $\pm$  1.0\\
    061222B & 3.355 &  40   &   22.4 $\pm$   1.83 &   1.59 $\pm$   0.36 &   0.635  $\pm$  2.486  &    29.1 $\pm$  21.1  &   1.296 $\pm$   0.565  &    46.7 $\pm$  15.7\\
    060927 & 5.6 & 22.5 &   11.3  $\pm$  0.68 &   2.7  $\pm$   0.17   & 0.339  $\pm$  0.462 &    125.4 $\pm$  60.5  &   0.919 $\pm$   0.378 &     72.0 $\pm$  17.6  \\
    060926 & 3.208 & 8 &      2.19 $\pm$   0.25  &  1.09  $\pm$  0.14  &  1.689  $\pm$  1.029 &     15.5 $\pm$  16.0 &    1.984 $\pm$   0.100  &     1.0 $\pm$   10.3 \\
    060908 & 1.884 &  19.3 &   28  $\pm$    1.11  &  3.03  $\pm$  0.25 &   0.134  $\pm$  0.614 &     115.2  $\pm$ 57.1 &    0.967 $\pm$   0.269 &    150.7 $\pm$ 112.4 \\
    060604 & 2.1357&  95  &    4.02  $\pm$  1.06  &  0.34  $\pm$  0.13  &  1.650 $\pm$   0.100 &    177.4  $\pm$  1.0  &   1.533 $\pm$   0.100  &    34.0  $\pm$  1.0\\
    060526 & 3.21  & 298.2 &  12.6  $\pm$  1.65  &  1.67  $\pm$  0.18  &  0.247 $\pm$   0.916  &    84.8 $\pm$  59.6 &    2.058  $\pm$  0.100  &    27.9  $\pm$  1.0 \\
    060512 & 0.4428 & 8.5  &   2.32  $\pm$  0.4  &   0.88 $\pm$   0.2  &  -0.163 $\pm$   2.503 &     22.9 $\pm$  11.1 &    1.011  $\pm$  1.752  &    23.0  $\pm$ 16.0 \\
    060510B & 4.9&     275.2 &   40.7 $\pm$    1.76  &   0.57  $\pm$   0.11  &   1.381  $\pm$   1.419  &    944.0 $\pm$    1.0   &   1.494  $\pm$   0.289  &    96.1 $\pm$    1.0 \\
    060218  &0.033&   2100 &   15.7  $\pm$  1.52  &  0.25  $\pm$  0.11 &   0.201 $\pm$   2.175 &     29.1 $\pm$  18.5  &   0.201  $\pm$  2.175   &   29.1 $\pm$  18.5\\
    060206 & 4.045 &  7.6  &   8.31  $\pm$  0.42  &  2.79  $\pm$  0.17 &   0.746  $\pm$  0.427  &    74.0  $\pm$ 17.0  &   1.165  $\pm$  0.325   &   78.1 $\pm$  25.6 \\
    060115  &3.53 &   139.6 &  17.1   $\pm$ 1.5  &   0.87   $\pm$ 0.12 &   1.279   $\pm$ 1.178  &   669.6  $\pm$  1.0  &   1.026  $\pm$  0.524   &   62.4 $\pm$  23.1 \\
    050525A  &0.606&  8.8  &   153   $\pm$  2.21 &   41.7   $\pm$ 0.94  &  0.584  $\pm$  0.144  &   109.0   $\pm$ 9.0  &   0.981   $\pm$ 0.118   &   82.3  $\pm$  3.7\\
    050416A & 0.653 &  2.5 &    3.67  $\pm$  0.37 &   4.88  $\pm$  0.48 &  -0.513  $\pm$  1.710  &    24.1 $\pm$   4.1 &    1.233  $\pm$  1.201  &    13.0  $\pm$ 10.3\\
    050406  &2.44 &   5.4  &   0.68   $\pm$ 0.14  &  0.36  $\pm$  0.1  &   0.054   $\pm$ 2.505  &    26.2  $\pm$ 18.6  &  -0.102  $\pm$  2.331   &   27.9  $\pm$ 10.7\\
    050318  &1.44  &  32  &    10.8   $\pm$ 0.77   & 3.16   $\pm$ 0.2  &   0.882   $\pm$ 0.459  &    70.3  $\pm$ 20.9  &   1.219  $\pm$  0.434  &    51.3  $\pm$ 11.2\\
  \hline
\end{longtable}
}


\end{document}